\documentclass[12pt]{iopart}
\usepackage{amsfonts}
\usepackage{amssymb}
\def\a{\alpha} 
\def\b{\beta}

\begin{document}

\title{An invariant joint alternative, by frames, to Einstein and Schroedinger equations}

\author{Shmuel Kaniel}

\address{Institute of Mathematics, Hebrew University of
   Jerusalem }
\ead{kaniel@math.huji.ac.il}
\begin{abstract}
The Hodge-de Rham Laplacean is an extension to forms of the wave equation.
A frame is a quartuple of 1-forms. The Hodge-de Rham Laplacean is modified to model it on the frame itself (not on the standard frame $dx$). This modified Laplacean is invariant.
The basic equation is: The modified Laplacean operating on the frame is equal to a source term times the frame. Kaniel and Itin (Il Nuovo Cimento vol 113B,N3,1998 )  analyzed the equation for steady state and spherically symmetric frame (General Relativity). They computed a closed solution for which the derived metric is Rosen's. This closed  solution is intrinsically different than Schwarzschild solution. Yet it passes the three classical experimental tests to the same accuracy.

The same basic equation is,also, the alternative to Schroedinger equation,
where the source term is the electromagnetic potential. The same quantization as in Schroedinger equation is attained.
The suggested system is hyperbolic, in contrast to the time dependent Schroedinger equation which is parabolic.
For time dependent frames the equation is quite complicated. Thus, the linearized equation is explicitly solved (Schroedinger equation is, already, linear). 

\end{abstract}
\section{Einstein and Schroedinger equations. The Hodge-de Rham Laplacean} 

A. Einstein\cite{1} in the theory of general relativity postulated that the world is a four dimensional Riemannian  manifold. Gravity is a field for which, in vacuum, the constitutive invariant equation is:
\begin{equation}\label{1}
R=0\,,
\end{equation}
where $R$ is the Ricci tensor. Furthermore, he postulated that massive bodies (treated as particles) move on geodesics. The postulates above enabled him to verify experimentally his theory. Later he suggested, for not empty space, the field equation to be  
\begin{equation}\label{2}
R-\lambda I=T\,,
\end{equation}
where $T$ is the energy-momentum tensor. This equation was not as successful  as the field equation in vacuum. 

On the other hand, the atomic structure is explained by Quantum theory. Its early  construct is Schroedinger equation. The time dependent equation is \cite{2}
\begin{equation}\label{3}
i\hbar\frac{\partial \psi}{\partial t}=-\frac{\hbar^2}{2\mu}\triangle \psi+\frac{Ze^2}{r}\psi\,,
\end{equation}
where $\psi$ is the wave function, $h$ is Planck constant, $\hbar=h/2\pi$,  $\mu$ is the mass of electron, $\triangle={\partial^2}/\partial x^2+ {\partial^2}/\partial y^2+{\partial^2}/\partial z^2$, $Z$ is an integer and $e$ is the charge of the electron. 
For the spherically symmetric case, the time independent equation is 
\begin{equation}\label{4}
-\frac{\hbar^2}{2\mu}\frac 1{r^2}\frac{d}{dr}\left(r^2\frac{d\psi}{dr}\right) -\frac{Ze^2}{r}\psi +\frac{\hbar^2}{2\mu}\frac 1{r^2}l(l+1)\psi=E\psi\,,
\end{equation}
The solution of (\ref{4}) is quantized 
 \begin{equation}\label{5a}
 E=E_n=-\frac{\mu Z^2e^4}{2\hbar n^2}\,.
 \end{equation}
 Eq. (\ref{3}) is equivalent to 
  \begin{equation}\label{5}
 \frac{\hbar^2}{2\mu}\left(\square(e^{-\alpha ct}\psi)+\frac 1{r^2}l(l+1)e^{-\alpha ct}\psi\right)={Ze^2}\frac 1{r}e^{-\alpha ct}\psi\,,
 \end{equation}
 where $\square =\frac 1{c^2}\frac {\partial^2}{\partial t^2}-\triangle$ is the  wave operator. $\alpha^2=-\frac{2\mu}{\hbar^2}E$. 

The domain of  Schroedinger equation is an abstarct Hilbert space. It is not invariant. 
In this article, it is suggested,  to replace (\ref{1}) and (\ref{2}) of General Relativity and (\ref{3}) of Quantum Mechanics by one invariant equation in a four dimensional space. 

The basic construct is the frame $\Phi$, as introduced by Cartan \cite{3}. 
A frame is a fourtuple of 1-forms $\Phi=(\Phi^0,\Phi^1,\Phi^2,\Phi^3)$
\begin{equation}\label{5x}
\Phi^r=\Phi^r_\beta dx^\beta\,,\qquad r,\beta=0,1,2,3\,,
 \end{equation}
 $\square$ is the restriction to functions of   Hodge-de Rham Laplacean on forms 
 \begin{equation}\label{6}
\square=d*d*+*d*d\,,
 \end{equation}
 where $d$ and $*$ are defined in the appendix. 
 
 Denote
  \begin{equation}\label{7}
\square_\Phi=d*_\Phi d*_\Phi+*_\Phi d*_\Phi d\,,
 \end{equation}
 $*_\Phi$ is modelled on the frame $\Phi$  as specified  in the appendix ($*$ is modelled on the Euclidean frame $dx$). Unlike $\square$, $\square_\Phi$ is an invariant operator. 
 
 The suggested equation is 
 \begin{equation}\label{8}
diag\left(\square_\Phi\Phi-\lambda(x)\Phi\right)=0\,,
 \end{equation}
 
 The metric for General Relativity as well as the wave function are completely determined by (\ref{8}). For the time dependent equation, the off diagonal terms, to be specified in the sequel, are completely determined by the diagonal. They do appear in $(\square_\Phi\Phi)^0_0$. 
 
 For General Relativity $\Phi$ is steady state and spherically symmetric. It satisfies $\Phi^\mu_\beta=0$, $\mu\ne\beta$. Thus the diagonal in (\ref{8}) can be omitted. The equation will be 
   \begin{equation}\label{9}
\square_\Phi\Phi-\lambda(x)\Phi=0\,,
 \end{equation}
 For Schroedinger equation $\Phi^0_j\ne 0$, $j=1,2,3$ and $\Phi^j_\mu=0$, $j=1,2,3$, $\mu=0,1,2,3$, $j\ne \mu$. 
 
 (\ref{8}) is complemented to a full invariant equation by the addition of $(\square_\Phi\Phi)^\mu_\beta$ for $\mu\ne \beta$. These terms are defined by $\Phi^\a_\a$ and $\Phi^0_j$. Adding them does not supply new information. 

Eq. (\ref{9}) is dealt with by Kaniel and Itin in \cite{4}. There, a closed solution to (\ref{9}) is computed. 
 The frame is 
 \begin{equation}\label{10}
\Phi^0=e^{-m/r}dx^0\,,\qquad \Phi^i=e^{m/r}dx^i\,.
 \end{equation}
 This frame yields the Rosen metric \cite{7}
 \begin{equation}\label{11}
ds^2=e^{-2m/r}dt^2+e^{2m/r}\left(dx^2+dy^2+dz^2\right)\,.
 \end{equation}
 The solution (\ref{10}) is essentially different from the Schawarzschild solution. The curvature is $$\frac{2m^2}{r^4}e^{-2m/r}\ne 0\,.$$
 The singularity is a point singularity unlike Schawarzschild radius. Neverless, black holes do exist. The two metrics are indistinguishable with respect to three classical experimental tests. 
 
 $\square_\Phi\Phi$ is  a very complicated object, cf \cite{5}. Consequently, let us compute it's linearization $L\square_\Phi\Phi$. It is invariant to the first order. Recall that A. Einstein had computed, first, the linearized equation \cite{1}. 
 
 For $\Phi$, to be exhibited, it will be shown that, in the terrestial coordinate system,
 \begin{equation}\label{12}
L\square_\Phi\Phi=\square\Phi\,.
 \end{equation}
Since 
 \begin{equation}\label{13}
(L*_\Phi) Ld  (L*_\Phi)  d=*d*d\,
 \end{equation}
 it is enough to prove that, for this particular frame 
 \begin{equation}\label{13x}
d (L*_\Phi) Ld  (L*_\Phi)\Phi=d*d(L*_\Phi)\Phi=d*d*\Phi\,,
  \end{equation}
or
  \begin{equation}\label{14}
d  (L*_\Phi)\Phi=d*\Phi\,.
  \end{equation}
 Then, when $\lambda(x)$ will be specified,  the linearization of (\ref{8}) 
  \begin{equation}\label{15}
diag\{ L\square_\Phi\Phi-\lambda(x)\Phi\}=0\,.
  \end{equation}
 will satisfy Schroedinger equation, which is, already, linear. 
 
 \section{The Linearized Equation}
 Consider a frame
 $$\Phi^\alpha_\beta=(I+\chi)^\alpha_\beta=\delta^\alpha_\beta+f^\alpha_\beta$$
 The $f^\alpha_\beta$ is assumed to  be small. The products of  $f^\alpha_\beta$ or their derivatives are neglected. 
 
 Let us compute, first, the linearization of  equation (\ref{9}). For General Relativity, $\lambda(x)=0$. The equation will be 
  \begin{equation}\label{15x}
 L\square_\Phi\Phi=0\,.
  \end{equation}
 Denote $x^0=ct$
  \begin{equation}\label{16}
 \Phi^0=[1-f(r)]dx^0\,,\qquad \Phi^j=[1+g(r)]dx^j
  \end{equation}
 To the first order,
 $$*_\Phi\Phi^0=\Phi^1\wedge\Phi^2\wedge\Phi^3=(1+3g)dx^1\wedge dx^2\wedge dx^3$$
 $$*\Phi^0=(1-f)dx^1\wedge dx^2\wedge dx^3$$
 $$Ld*_\Phi\Phi^0=d*\Phi^0=0$$
 Consequently, by (\ref{16}),
  \begin{equation}\label{17}
 L\square_\Phi\Phi^0=\square\Phi^0=0
  \end{equation}
  Thus, 
   \begin{equation}\label{18}
\Phi^0=(1-\frac mr)dx^0
  \end{equation}
  Now
  $$*_\Phi\Phi^1=\Phi^0\wedge\Phi^2\wedge\Phi^3=(1+2g-f)dx^0\wedge dx^2\wedge dx^3$$
  $$*\Phi^1=(1+g)dx^0\wedge dx^2\wedge dx^3$$
  If $g=f$ then $*_\Phi\Phi^1=*\Phi^1$ and $\square_\Phi\Phi^1=\square\Phi^1$
  Thus
    \begin{equation}\label{19}
\Phi^j=(1+\frac mr)dx^j
  \end{equation}
  The line element will be 
  $$ds^2=(1-\frac {2m}r){dx^0}^2+(1+\frac {2m}r)dr^2$$
  The linearized Einstein element.
  
  The essential equation $f=g$ results from the definition of $\square_\Phi\Phi$. $\square$ by itself is not enough. 
  
  Ansatz for the Schroedinger frame. Two versions exist, real and complex. The real one is  
  \begin{eqnarray}\label{22x}
  \Phi^0&=&-exp(-\alpha x^0) f(x)dx^0\nonumber\\
   \Phi^j&=&exp(-\alpha x^0) \left(g(x)\frac{x^j}rdx^0+f(x)dx^j\right)
\end{eqnarray}
Let us compute $L\square_\Phi\Phi$. For simplicity, omit the $L$. 
When (\ref{22x}) is substituted into (\ref{8}) the common factor $exp(-\alpha x^0)$ may be cancelled. The equation will be time independent. 

\begin{eqnarray} \label{22}
&&*_\Phi\Phi^0=\Phi^1\wedge\Phi^2\wedge\Phi^3=dx^1\wedge dx^2\wedge dx^3+e^{-\alpha x^0} \{3fdx^1\wedge dx^2\wedge dx^3+\nonumber\\
&&\qquad g\frac 1r[x^1dx^0\wedge dx^2\wedge dx^3+x^2dx^0\wedge dx^3\wedge dx^1+x^3dx^0\wedge dx^1\wedge dx^2]\}
\end{eqnarray}
and 
\begin{eqnarray} \label{23}
&&*_\Phi\Phi^1=\Phi^0\wedge\Phi^2\wedge\Phi^3=e^{-\alpha x^0} fdx^0\wedge dx^2\wedge dx^3=*\Phi^1
\end{eqnarray}
(\ref{22}) and (\ref{23}) yield
\begin{eqnarray} \label{24}
&&d*_\Phi\Phi^0-d*\Phi^0=e^{-\alpha x^0}\left[4\a f-\left(g_{,i}\frac{x^i}r+2g\frac 1r\right)\right]dx^0\wedge dx^1\wedge dx^2\wedge dx^3\,.\nonumber\\ &&
\end{eqnarray}
Consequently, if
\begin{eqnarray} \label{25}
&&4\a f-\left(g_{,i}\frac{x^i}r+2g\frac 1r\right)=0\,.
\end{eqnarray}
Then, by (\ref{16},\ref{17},\ref{23},\ref{24}) and (\ref{25})
 \begin{equation}\label{26}
L\square_\Phi\Phi=\square\Phi\,.
  \end{equation}
  (\ref{25}) can be explicitly solved. The solution of Schroedinger equation and of (\ref{8}) are 
  $$\psi=f=F(r)Y^{l,m}(\theta,\phi)\,,$$
  where $Y^{l,m}$ are the spherical harmonics, c.f. \cite{2},\cite{6}.  
  Take, also, $g=G(r)Y^{l,m}$. Since, for any $R(r)$
  \begin{equation}\label{27}
  \triangle \left(R(r)Y^{l,m}\right)=\triangle R(r)Y^{l,m}+R(r)\triangle Y^{l,m}
  \end{equation}
  it follows that 
  $$R(r)_{,j}Y^{l,m}_{,j}=0$$
  Thus
  \begin{equation}\label{28}
  4\a F-\left(G_{,r}+2G\frac 1r\right)=0
  \end{equation}
  and
   \begin{equation}\label{29}
  G=4\a r^{-2}\int r^2Fdr
  \end{equation}
  By, (\ref{29}), 
  $$\square e^{-\a x^0}g\frac {x^i}r =\square e^{-\a x^0} GY^{l,m}x^i\frac 1r$$
  is easily computed. 
  
  For the complex version, eq. (\ref{5}) is modified to be 
  \begin{equation}\label{33}
 \frac{\hbar^2}{2\mu}\left(\square(e^{i\alpha ct}\psi)+\frac 1{r^2}l(l+1)e^{i\alpha ct}\psi\right)=-{Ze^2}\frac 1{r}e^{i\alpha ct}\psi\,,
 \end{equation}
 For that the factor $i$ will stand in front of $g$ and $G$. 
 \appendix
  \section  {}
  A frame $\Phi$ yields the metric $g$ by 
  $$g_{\mu\nu}=\eta_{ab}\Phi^a_\mu\Phi^b_\nu$$
  where $\eta_{ab}=diag(-1,1,1,1)$ the Lorentzian metric tensor. 

  The operator $d$ acts on exterior forms 
  \begin{equation}\label{A1}
  d\left(f(x)dx^{\a_1}\wedge\cdots\wedge dx^{\a_k}\right)=f(x)_{,\a} dx^\a\wedge dx^{\a_1}\wedge\cdots\wedge dx^{\a_k}
  \end{equation}
$*$ is the hyperbolic star. Denote $dx^0=cdt$. Consequently
    \begin{equation}\label{A2}
  *\left(dx^{\a_1}\wedge\cdots\wedge dx^{\a_k}\right)=(-1)^ldx^{\b_1}\wedge\cdots\wedge dx^{\b_{n-k}}
  \end{equation}
  where $\a_1,\cdots,\a_k,\b_1,\cdots,\b_{n-k}$ is an even permutation of $(0,1,2,3)$. $l=1$ if zero is one of the $\b_i$, $l=0$ otherwise. 

The Hodge-de Rham Laplacian is defined by
  \begin{equation}\label{A3}
  \square=d*d*+*d*d
   \end{equation}
   On functions and 1-forms 
    \begin{equation}\label{A4}
    \square f=\frac{\partial^2 f}{\partial x_0^2}-\triangle f;\qquad 
     \square (fdx^\a)=\square f dx^\a
       \end{equation}
       \begin{equation}\label{A5}
       \square_\Phi=d*_\Phi d*_\Phi+*_\Phi d d*_\Phi d
        \end{equation}
where   
          \begin{equation}\label{A6}
  *_\Phi\left(\Phi^{\a_1}\wedge\cdots\wedge \Phi^{\a_k}\right)=(-1)^l\Phi^{\b_1}\wedge\cdots\wedge \Phi^{\b_{n-k}}
  \end{equation}

  \end{document}